\documentclass[reqno, oneside]{amsart}
\usepackage{style}

\newcounter{theorem}
\newenvironment{defi}[1]{\par\addvspace{0.5cm}
\begin{sloppypar}\refstepcounter{theorem}%
{\bf #1 \thetheorem.}\rm{}}{\end{sloppypar}}

\begin{document}
\title[Weighted Paley-Wiener spaces]{An inverse problem for weighted \\ Paley-Wiener spaces}
\author{R.~V.~Bessonov, \;\; R.~V.~Romanov}

\address{St.Petersburg State University ({\normalfont \hbox{29b}, 14th Line V.O., St.Petersburg,  199178, Russia}) and St.Petersburg Department of Steklov Mathematical Institute of Russian Academy of Science ({\normalfont 27, Fon\-tan\-ka, St.Petersburg, 191023, Russia})}
\email{bessonov@pdmi.ras.ru}

\address{St.Petersburg State University ({\normalfont  7/9 Universitetskaya nab., St.Petersburg, 199034
Russia})}
\email{morovom@gmail.com}

\thanks{The work is supported by Russian Science Foundation Grant 14-21-00035 (Theorem 1) and by Russian Science Foundation Grant 14-41-00010 (Theorem 2).}

\begin{abstract}
Let $\mu$ be a measure on the real line $\R$ such that $\int_{\R}\frac{d\mu(t)}{1+t^2} < \infty$ and let $a>0$. Assume that the norms $\|f\|_{L^2(\R)}$ and $\|f\|_{L^2(\mu)}$ are comparable for functions $f$ in the Paley-Wiener space $\pw_{a}$ and that $\pw_a$ is dense in $L^2(\mu)$. We  reconstruct the canonical Hamiltonian system $JX' = z \H X$ such that $\mu$ is the spectral measure for this system.
\end{abstract}

\maketitle

\section{Introduction}\label{s1}
\noindent 
The standard procedure in the inverse spectral theory of ordinary differential operators \cite{LSbook}, \cite{Mar06} is based on construction of transformation operators, that is, operators mapping eigensolutions of an unperturbed differential equation into those of the perturbed one. The procedure for recovery of the transformation operators, be it the Gelfand-Levitan-Marchenko formalism or the Krein equation, is via solving compact integral equations for the corresponding kernels. In the context of operators with rough coefficients, the strongest results of this approach so far are the inverse spectral theory for Dirac operators with $ L^1 $-potentials constructed in \cite{AlbeverioGrinivMykytyuk} and the theory of Schroedinger operators with $ W_2^{ -1 } $ potentials \cite{HrynivMykytyuk2003, HrynivMykytyuk2004}. A natural limitation for this kind of argument is that the operators in the integral equations have to be (at least) compact perturbations of unity.

\medskip

In the present paper we pursue a different statement of the problem. Rather than imposing conditions on spectral data/coefficients of differential operator, which imply the existence of transformation operators, we characterize the spectral measures for which bounded transformators exist (Theorem \ref{t4}). Since the transformators obtained generally have no good kernel properties, their existence per se does not imply a solution of the recovery problem, for no analogue of the standard formulae expressing the potential via values of the kernel is possible. Our main result (Theorem \ref{t1}) is a recovery procedure for an operator from its spectral measure which does not use such formulae.  All these are done in a model situation of Dirac-like canonical systems which we describe now.

\subsection{Existence of transformation operators} 
Consider a canonical Hamiltonian system of order two on a finite interval $[0, \ell]$,
\begin{equation}\label{eq1}
JX'(r,z) = z \H(r) X(r,z), \qquad 0<r<\ell, \quad z\in \C.
\end{equation}
Here the Hamiltonian $\H$ is a mapping from $[0, \ell]$ to the set of $2\times2$ non-negative matrices with real entries, $\trace\H \in L^1[0,\ell]$, the derivative of $X: [0,\ell]\times \C \to \C^2$ is taken in $r$, 
$
J = 
\left(
\begin{smallmatrix}
0 &-1 \\
1 &\,\,\,0 
\end{smallmatrix}
\right)
$. Define $M$ to be the fundamental matrix of the system \eqref{eq1}, that is,
\begin{equation}\label{eq66}
JM' = z \H M, \;\; 
M(r,z) = 
\begin{pmatrix}
\Theta_\H^+(r,z) &\Phi_\H^+(r,z)\\
\Theta_\H^-(r,z) &\Phi_\H^-(r,z)
\end{pmatrix}, 
\;\;
M(0,z) = 
\begin{pmatrix}
1 &0\\
0 &1
\end{pmatrix}. 
\end{equation}

Let us recall the notions of a transformation operator and a spectral measure.   Define $\Theta_{\H} = \Theta_{\H}(r,z)$ to be the first column of the matrix $ M $, and let $\H_0 = \idm$. Then $\Theta_{\H_0}= \cs$.  The transformation operator $S_\H$ is by definition a linear map of appropriate function spaces such that 
\begin{equation}\label{SH}
S_\H: \Theta_{\H_0}(\cdot, z) \mapsto \Theta_{\H}(\cdot, z), \qquad z \in \C.
\end{equation}
This formula correctly defines $ S_\H $ on the linear span of $ \{ \Theta_{\H_0}(\cdot, z) \}_{ z \in \C } $ which is a dense set in  $ L^2( [0, a] , \mathbb C^2) $. For $r \in[0,\ell]$ define the Hilbert space
\begin{gather}\label{Hr}
L^2(\H, r) = \Bigl\{X \colon [0,r] \to \C^2 \colon \int_{0}^{r}\bigr\langle\H(s)X(s), X(s)\bigr\rangle_{\C^2} \,ds < \infty\Bigr\}\Big/{\mathcal Ker}\,\H,\\
{\mathcal Ker}\,\H = \Bigl\{X\colon \;\H(t) X(t) = 0 \mbox{ for almost all } t\in [0,r]\Bigl\}. \notag
\end{gather} 
The Weyl function, $\Phi_\H^-(\ell, \cdot)/\Theta_\H^-(\ell, \cdot)$, has positive imaginary part in the upper half-plane of the complex plane $\C$, hence
there is a measure $\mu \ge 0$ such that $\int_{\R}\frac{d\mu(t)}{1+t^2} < \infty$ and
\begin{equation}\label{eq47}
\frac{\Phi_{\H}^{-}(\ell, z)}{\Theta_{\H}^{-}(\ell, z)} = \frac{1}{\pi}\int_{\R}\left(\frac{1}{t-z} - \frac{t}{1+t^2}\right)\, d\mu(t) + bz + c, \qquad \Im z > 0 ,
\end{equation}
for some $b\ge 0$, $c \in \R$. We will refer to $\mu$ as the {\it principal} spectral measure, see Section \ref{s2} for details. By the inverse problem we mean restoring $ \H $ from the principal spectral measure. 

\medskip

Fix $a>0$ and denote by $\pw_a$ the Paley-Wiener space of functions in $L^2(\R)$ whose Fourier transform vanishes outside $[-a,a]$. We consider measures $\mu$ having the following properties:
\begin{itemize}
\item[(1)] $\mu$ is a measure on $\R$ such that $\mu(\{0\}) > 0$ and $\int_{\R}\frac{d\mu(t)}{t^2+1} < \infty$; 
\item[(2)] the norms $\|f\|_{L^2(\mu)}$ and $\|f\|_{L^2(\R)}$ are comparable for 
$f \in \pw_a$;
\item[(3)] the space $\pw_a$ is dense in $L^2(\mu)$.
\end{itemize}

\begin{Thm}\label{t4}
The following assertions are equivalent:
\begin{itemize}
	\item[$(a)$] $\mu$ satisfies assumptions $(1) -(3)$;
	\item[$(b)$] $\mu$ is the principal spectral measure  for a Hamiltonian $\H$ on $[0,\ell]$ such that $S_\H$ extends to a bounded and boundedly invertible operator from $L^2(\H_0,[0, a])$ onto $L^2(\H, [0, \ell])$.  
\end{itemize}
\end{Thm}

The class of measures satisfying (1)--(3) contains spectral measures of canonical systems corresponding to Dirac operators $L_Q: X \mapsto JX' + QX$ with selfadjoint potentials $Q \in L^1[0, \ell]$ (the reduction of the Dirac opeartor to a canonical system can be found e. g. in \cite{Romanov}). The Dirac operator on the interval $ [0 , 1 ] $ with an $ L_1 $-potential in the standard form has the Neumann-Dirichlet and Neumann spectra, $ \lambda_n $, $ \mu_n $,  such that $ \{ \lambda_n - \pi ( n + 1/2 ) \} $, $ \{ \mu_n - \pi n \} $ resp., are the Fourier coefficients of $ L^1 $-functions \cite{AlbeverioGrinivMykytyuk}. In the classical theory, the inverse problem for these data is solved via the Krein equation for an appropriate kernel whose existence is derived from the smallness of the reminder in the spectral asymptotics. The existence of transformation operators for measures in Theorem \ref{t4} does not follow from this type of argument. For instance, an inverse problem with a measure $ \mu $ supported on the set $ \{ \pi n + d_n \}_{ n \in \mathbb Z } $ with $ \sup_n |d_n | < \pi/4 $, $ \inf_n |d_n | > 0 $, apparently cannot be handled via the Gelfand-Levitan-Marchenko-Krein construction, while such measures do exist among those satisfying conditions of Theorems \ref{t4} and \ref{t1},  by the Kadec $1/4$--theorem~\cite{Levinlectures}. Instead of integral equations for kernels, we apply a general inverse spectral theory by L. de Branges \cite{dbbook}. Notice that we use the de Branges theory to construct the algorithm, rather than in a  uniqueness question \cite{EckhardtKostenkoTeschl}. A motivation to study inverse problems in the class of measures $ \mu $ under consideration can be found in paper \cite{Remling} by C.~Remling.

\medskip

It is an open problem to find a tractable description of Hamiltonians $\H $ for which the condition $ (b) $ of Theorem \ref{t4} is satisfied. A simple explicit necessary condition can be obtained by successive differentiation of (\ref{SH}) at $ z = 0 $, see Proposition \ref{p2} for the case of diagonal Hamiltonians. Towards this problem, we would like to make the following observation. It is motivated by the fact that the Hamiltonians corresponding to Dirac operators \cite{Den06,Romanov} discussed above are bounded and boundedly invertible in the sense that 
$\H, \H^{-1} \in L^\infty[0,\ell]$. In Proposition \ref{separated} we show that the latter property alone does not imply that the corresponding space $L^2(\mu)$ is norm equivalent to a Paley--Wiener space $\pw_a$, that is, we construct a bounded and boundedly invertible Hamiltonian such that the property $(2)$ does not hold for the corresponding principal spectral measure.  

\medskip

\subsection{Recovery algorithm} 
The recovery algorithm for Hamiltonians of canonical systems whose principal spectral measures have properties $(1)-(3)$ could be summarized as follows. Take a number $s \in [0, a]$. By property $(2)$, the truncated Toeplitz operator, $ T_{\mu,s} $, \cite{BBK,Sar07}
\begin{equation}\label{eq75}
(T_{\mu,s} f, g) = \int_\R f\bar g \,d\mu, \qquad f,g \in \pw_s,
\end{equation}
is correctly defined, positive, and invertible on the Paley-Wiener space $\pw_s$. Define functions in $L^2(\mu)$ by
\begin{equation}\label{eq65}
\begin{aligned}
&G_{1,s}: t \mapsto \left(T_{\mu,s}^{-1}\frac{\sin sx}{x}, \frac{\sin s(x - t)}{\pi(x - t)}\right)_{L^2(\R)}, \\
&G_{2, a}: t \mapsto
\begin{cases}
 \frac{\hat{K}_{\mu}(0) + c}{\mu(\{0\})}- G'_{1,a}(0)\mu(\{0\}),       &t=0, \\
 \frac{1}{t}\left(\frac{\pi}{t \mu(\{t\}) G'_{1,a}(t)} - 1 \right),                &t \in\supp\mu\setminus\{0\}, 
\end{cases}\\
&G_{2,s}: t \mapsto\left(G_{2,a}, \; T_{\mu,s}^{-1}\frac{\sin s(x-t)}{\pi(x - t)}\right)_{L^2(\mu)},
\end{aligned}
\end{equation}
where $c$ is the constant from formula \eqref{eq47}, $\hat{K}_{\mu}(0) = \frac{1}{\pi}\int_{\R\setminus\{0\}}\frac{d\mu(t)}{t(1+t^2)}$, and
\begin{equation}\notag
G'_{1,a}(t) = \left(T_{\mu,a}^{-1}\frac{\sin ax}{x}, \frac{\sin a(x-t) - (x-t)\cos a(x-t)}{\pi (x-t)^2}\right)_{L^2(\R)}.
\end{equation}
Consider the strictly increasing bijection $\zeta: s \mapsto \frac{1}{2}G_{1,s}(0) + \frac{1}{2\pi}\|G_{2,s}\|_{L^2(\mu)}^{2}$ mapping the interval $[0, a]$ onto $[0, \ell]$, where $\ell = \zeta(a)$. 
Let $\tau: [0,\ell] \to [0, a]$ be the inverse function: $\zeta(\tau(r)) = r$ for all $r \in [0, \ell]$. The functions 
\begin{equation}\label{eq111}
g_1 : r \mapsto G_{1, \tau(r)}(0), \qquad g_2 = 2r - g_1, \qquad g: r \mapsto \frac{1}{\pi}\bigl(G_{1, \tau(r)}, G_{2, \tau(r)}\bigr)_{L^2(\mu)}, 
\end{equation}
turn out to be absolutely continuous on $[0, \ell]$. We denote by $g'_1$, $g'_2$, $g'$ their derivatives in $L^1[0,\ell]$.
\begin{Thm}\label{t1} Assume that $\mu$ is a measure with properties $(1) - (3)$. Then $\mu$ is the principal spectral measure corresponding to the Hamiltonian $\H$ on $[0, \ell]$ given by
\begin{equation}\label{eq48}
\H = 
\begin{pmatrix}
g'_1 & g' \\
g' & g'_2
\end{pmatrix}.
\end{equation}
\end{Thm}

\medskip

\noindent {\bf Acknowledgement.} The authors are grateful to N.~G.~Makarov who suggested to use truncated Toeplitz operators in the setting of canonical systems. In particular, he noticed that $ T_{ \mu , a } $ takes reproducing kernels of the perturbed problem into the Paley-Wiener ones (see Lemma 4.2 in the
current text) and proposed to exploit this observation as a replacement of Gelfand-Levitan-Marchenko-type theories. The hospitality of California Institute of Technology during the visit of the first author in 2012 is greatly appreciated.

\section{Some known facts}\label{s2}
In this section we recall some facts from the direct spectral theory of canonical Hamiltonian systems and de Branges theory of Hilbert spaces of entire functions. All information we need can be found in de Branges book \cite{dbbook} or Lecture notes \cite{Romanov}. 

\subsection{Weyl-Titchmarsh transform}\label{wt} A mapping $\H$ from $[0,\ell]$ to the set of all non-negative matrices with real entries is called regular Hamiltonian if $\trace \H \in L^1[0, \ell]$ and there is no interval $I \subset [0,\ell]$ such that $\H = 0$ on $I$. 
We will say that the Hamiltonian $\H$ compatible at $0$ (correspondingly, at $\ell$) if there is no interval $I = [0, \eps)$ (correspondingly, $I = (\ell-\eps, \ell]$) such that $\H$ coincides with scalar multiple of the rank-one operator $\left\langle \cdot, \zo\right\rangle_{\C^2}\!\zo$ almost everywhere on $I$. The compatibility condition at $0$ guarantees that the solution of Cauchy problem 
\begin{equation}\label{eq73}
JX'(r, z) = z\H(r) X(r, z), \quad X(0, z) = \oz
\end{equation}
is non-constant on any open subinterval of $[0, \ell]$.
An interval $I \subset [0, \ell]$ is called indivisible for $\H$ if there exists a vector $e \in \R^2$ such that $\H$ coincides with the rank-one operator $f \mapsto \langle f,e\rangle_{\C^2}e$ almost everywhere on $I$, and $I$ is the maximal interval (with respect to inclusion) having this property. 

\medskip

Let $\H$ be a regular Hamiltonian on $[0,\ell]$, 
and $\Theta_\H$ be the solution of the Cauchy problem \eqref{eq73}.
Denote the components of $\Theta_{\H}$ in the standard basis of~$\C^2$ by $\Theta_{\H}^{\pm}$. Remark that $\Theta_{\H}$ coincides with the first column of the fundamental matrix solution \eqref{eq66}. Assume that $\H$ is compatible at $0$.  Then for every $r \in (0, \ell]$ the function 
\begin{equation}\label{eq82}
E_{\H, r}: z \mapsto \Theta_{\H}^{+}(r, z) + i \Theta_{\H}^{-}(r, z)
\end{equation}
is an entire function of order at most~$1$ and of finite exponential type; it does not vanish on the real line. Moreover, $E_{\H,r}$  belongs to the Hermite-Biehler class, that is, $|E_{\H,r}(\bar z)| < |E_{\H,r}(z)|$ for all $z$ in the upper half-plane $\C_+=\{z \in \C:\; \Im z > 0\}$. Each function $E$ from the Hermite-Biehler class determines a Hilbert space of entire functions (de Branges space),
\begin{equation}\label{eq44}
\B(E) = \Bigl\{entire \; f: \; \frac{f}{E} \in H^2_+, \;\; \frac{f^*}{E} \in H^2_+\Bigr\},
\end{equation}
equipped with the inner product $(f,g)_{\B(E)} = (f/E,\,g/E)_{L^2(\R)}$. In the above formula $H^2_+$ stands for the classical Hardy space in $\C_+$, and we denote $f^*: z \mapsto \ov{f(\bar z)}$. 

Denote by $\mathfrak{I}(\H)$ be the set of indivisible intervals of the Hamiltonian $ \H $. For $r \in[0,\ell]$ consider the Hilbert space  
\begin{equation}\label{eq49}
H_r = \bigr\{X \in L^2(\H, r): X = x_I \mbox{ on } I\cap [ 0 , r] , I\in\mathfrak{I}(\H), \;x_I \in \C^2\Bigr\}.
\end{equation}
The Weyl-Titchmarsh transform,
\begin{equation}\label{eq77}
\mathcal W_r: X \mapsto \frac{1}{\sqrt{\pi}}\int_{0}^{r} \bigr\langle \H(t)X(t), \Theta_{\H}(t,\bar z) \bigr\rangle_{\C^2} \,dt, \qquad z \in \C,
\end{equation}
maps the space $H_r$ unitarily onto $\B(E_r)$. Moreover, the operator $\mathcal W_r$ takes solutions $\Theta_{\H}( \cdot ,\bar w)$ on the interval $[0, r]$ into the reproducing kernels of the de Branges space $\B(E_{\H,r})$, 
\begin{equation}\label{eq43}
\mathcal W_r \Theta_{\H}(\cdot,\bar w) = \sqrt{\pi}\cdot k^r_{w}, \quad
k^r_w(z) = \frac{1}{\pi}
\frac{\Theta_{\H,r}^{+}(z)\Theta_{\H,r}^{-}(\bar w) - \Theta_{\H,r}^{-}(z)\Theta_{\H,r}^{+}(\bar w)}{z - \bar w}.
\end{equation}
Now take $r=\ell$. Given a mapping $X: [0,\ell] \to \C^2$, denote by $X^{\pm}$ its components in the standard basis of $\C^2$: $X^+ = \langle X, \oz\rangle_{\C^2}$, $X^- = \langle X, \zo\rangle_{\C^2}$. 
The principal spectral measure $\mu$ defined in (\ref{eq43}) coincides with the measure
\begin{equation}\label{eq72}
\mu = \sum_{x_k \in \Lambda} \frac{\delta_{x_k}}{\|k_{x_k}^{\ell}\|_{\B(E_{\H,\ell})}^{2}}, \qquad 
\Lambda = \{x \in \R: \Theta_{\H}^{-}(\ell, x) = 0\},
\end{equation}
where $\delta_x$ denotes the Dirac measure concentrated at $x \in \R$. The restriction
\begin{equation}\label{eq114}
\Pi_\mu: f \mapsto f|_{\supp\mu}
\end{equation}
is a unitary operator from $\B(E_{\H, \ell})$ onto $L^2(\mu)$, and thus $\Pi_\mu \mathcal W_\ell  $ is an isomorphism between $ H_\ell $ and $ L^2(\mu)$. The term spectral applied to the measure $ \mu $ refers to the fact that $\Pi_\mu \mathcal W_\ell  $ diagonalizes a selfadjoint differential operator, $ L_\H $,  defined in $ H_\ell $ by 
\begin{equation}\label{eq41}
L_{\H}: X \mapsto Y, \quad JX' = \H Y,
\end{equation} 
on a natural domain corresponding to the boundary condition $X^-( 0 ) = X^-(\ell) = 0$, see \cite{HWdeSnoo}. To spell this out, $\Pi_\mu \mathcal W_\ell L_{\H} = M_x \Pi_\mu \mathcal W_{\ell}$, where 
$M_x$ is the operator of multiplication by the independent variable in $L^2(\mu)$.  

\subsection{Hilbert spaces of entire functions}
The following result is Theorem 23 in \cite{dbbook}.
\begin{ThmA}[de Branges]\label{A1}
For every Hermite-Biehler function $E$ the space $\B(E)$ has the following properties:
\begin{itemize}
\item[$(A_1)$] whenever $f$ is in the space and has a nonreal zero $w$, the function $\frac{z - \bar w}{z - w}f$ is in
the space and has the same norm as $f$;
\item[$(A_2)$] for every nonreal number $w$, the evaluation functional $f \mapsto f(w)$ is continuous;
\item[$(A_3)$] the function $f^*$ belongs to the space whenever $f$ belongs to the space and it always has the same norm as $f$.
\end{itemize}
Conversely, for every Hilbert space of entire functions $\mathcal{B}$ satisfying $(A_1)$-$(A_3)$ there exists a Hermite-Biehler function $E$ such that $\mathcal B = \B(E)$. 
\end{ThmA}
Next theorem is a simple corollary of Theorem 35 in \cite{dbbook}.
\begin{ThmA}[de Branges]\label{A2}
If two subspaces of a de Branges space are de Branges spaces themselves then one of them contains the other, provided that the corresponding Hermite-Biehler functions have no real zeroes. 
\end{ThmA}
An Hermite-Biehler function $E$ is called regular if $\int_{\R}\frac{1}{1+t^2}\frac{dt}{|E(t)|^2} < \infty$. As is easy to check, the function $E$ is regular if and only if for every $w \in \C$ and $F \in \B(E)$ we have $\frac{F - F(w)}{z - w} \in \B(E)$.    
\begin{ThmA}[de Branges]\label{A3}
For every regular Hamiltonian $\H$ on an interval $[0,\ell]$ compatible at $0$ its Her\-mite-Biehler function $E_{\H,\ell}$ in \eqref{eq82} is regular and has no real zeroes. Conversely, to every regular Hermite-Biehler function $E$ without real zeroes, $E(0) = 1$, and every $c>0$ there exist a unique $\ell > 0$ and a Hamiltonian $\H$ on $[0,\ell]$ compatible at $0$ such that $\trace \H = c$ almost everywhere on $[0,\ell]$, and~$E = E_{\H, \ell}$.
\end{ThmA}
It will be convenient for us to use Theorem \ref{A3} with $c = 2$, because we have $\trace \H_0 = 2$ for the ``free'' Hamiltonian $\H_0 = \idm$.

The following theorem is a corollary of the previous one and the lattice property of de Branges spaces.

\begin{ThmA}[de Branges]\label{A4}
Let $ E $ be a regular Hermite--Biehler function without real zeroes subject to the condition $ E ( 0 ) = 1$, and let $ \H $ be the Hamiltonian corresponding to this function in the sense of Theorem \ref{A3}. The set of subspaces of $ \B ( E ) $ which are de Branges spaces themself (with the norm inherited from $ \B ( E ) $) coincides with the collection $ \{ \Ran \mathcal W_r \} $ when $ r $ ranges over the complement, $ \mathcal M $, of inner points of indivisible intervals in $ [ 0 , \ell ] $.  
\end{ThmA} 

The collection $ \{ \Ran \mathcal W_r \}_{r \in \mathcal M } $ is called the de Branges chain corresponding to the function $ E $. 

\medskip

\section{Proof of Theorem \ref{t4}}

We will need two standard results from the theory of de Branges spaces. Their proofs are included for completeness.
We write $H_1  \sets H_2$ for Hilbert spaces $H_1$, $H_2$ if $H_1$ coincides with $H_2$ as a set and the norms in $H_1$ and $H_2$ are equivalent. Hermite-Biehler functions $E$ such that $\B(E) \sets \pw_a$ for some $a>0$ are described in Theorem 4 of \cite{LS}, see also \cite{Volberg}. 
\begin{Lem}\label{l10}
Let $\H$ be a regular Hamiltonian on an interval $[0, \ell]$ compatible at~$0$, and let $\B(E_{\H,r})$, $r \in [0, \ell]$, be its chain of de Branges subspaces. Assume that $\B(E_{\H,\ell}) \sets \pw_a$. Then $\H$ has no indivisible intervals and there exists an increasing bijection $\xi: [0, a] \to [0, \ell]$ such that 
$\B(E_{\H,\xi(s)}) \sets \pw_s$ for all $s \in [0,a]$.
\end{Lem}
\beginpf Denote $E_{\H,r} = E_r$. For every $r  \in (0,\ell )$ not being an inner point of an indivisible interval for $ \H $ define the Hilbert space
$$
\mathcal C_r = \Bigl(\B ( E_r ) , \;\; (\,\cdot\,, \,\cdot\,)_{\pw_a }\Bigr).
$$
 It is easy to check that $\mathcal C_r $ has properties 
$(A_1)-(A_3)$ from Theorem \ref{A1}. Hence $\mathcal C_r$ is a de Branges subspace of $\pw_a$. It follows from Theorem \ref{A4} that $\mathcal C_r$ coincides (as a Hilbert space) with a Paley-Wiener space $\pw_s$ for some $s \in [0,a]$. This immediately implies that $\H$ has no indivisible intervals, for otherwise there would exist $r,t\in [0,\ell]$ such that $\mathcal C_r$ is a subspace in  $\mathcal C_t$ of codimension $1$ which is impossible for two Paley-Wiener spaces. It follows that  the correspondence $r \mapsto s$ defines a monotone injection from $[0 , \ell] $ to $[0 , a] $. The lemma will be proved if we show that the range of this injection coincides with $ [  0 , a ] $. Indeed for any $s \in [0 , a]$ the space   
$$
\mathcal B_s = \Bigl(\pw_s, \;\; (\,\cdot\,, \,\cdot\,)_{\B(E_{\ell})}\Bigr),
$$
is a de Branges subspace of $\B(E_{\ell})$, again by Theorem \ref{A1}. This subspace coincides with $\mathcal C_t$ for some $t$, and it is clear that this $t$ goes to $s$ under the injection. \qed  

\medskip

\noindent {\bf Remark.} With some effort, Lemma \ref{l10} admits the following generalization: for every pair of regular Hamiltonians $\H_1$, $\H_2$ on $[0, \ell]$ which are compatible at $0$ and satisfy the relation $\B(E_{\H_1, \ell})\sets\B(E_{\H_2, \ell})$, there exists an increasing bijection $\xi: [0, a] \to [0, a]$ such that $\B(E_{\H_1,r}) \sets \B(E_{\H_2, \xi(r)})$ for all $r \in [0,\ell]$.   

\medskip

\noindent {\bf Remark.} The function $ \xi $ from Lemma \ref{l10} is obviously continuous. It is not known however if  this function $\xi$  is absolutely continuous on the interval $[0, a]$. The absolute continuity of $\xi$ is equivalent to the fact that the Hamiltonian $\H$ has rank two almost everywhere on $[0, \ell]$. Krein's formula for exponential type (see \cite{Romanov}) can be written in the form $s = \int_{0}^{\xi(s)}\sqrt{\det H(t)}\,dt$. As is easy to see from this formula, the inverse mapping $\tau$ to $\xi$ is an absolutely continuous function on $[0, \ell]$ and the Hamiltonian $\H$ cannot have rank one on a subinterval of $[0,\ell]$. 

\medskip

\begin{Lem}\label{l3}
Let $\H$ be a regular Hamiltonian on $[0,\ell]$ compatible at $0$, $\ell$, and let $\mu$ be a positive discrete measure on~$\R$ 
such that $\mu(\{0\}) > 0$ and $\int_\R\frac{d\mu(t)}{1+t^2} < \infty$. Assume that the restriction $\Pi_\mu: f \mapsto f|_{\supp\mu}$ from 
$\B(E_{\H,\ell})$ to $L^2(\mu)$ is a unitary operator. Then $\mu$ is the corresponding principal spectral measure, that is, it satisfies~\eqref{eq47}.
\end{Lem}
\beginpf Consider the function $k = \Pi_{\mu}^{-1}\chi_{\{0\}}$ in $\B(E_{\H,\ell})$, where $\chi_{\{0\}}$ is the indicator of the singleton $\{0\}$. Since $\Pi_\mu$ is isometric, we have
$$
(f,k)_{\B(E_{\H,\ell})} = (\Pi_\mu f, \chi_{\{0\}})_{L^2(\mu)} = f(0)\mu(\{0\})
$$
for every $f \in \B(E_{\H,\ell})$. It follows that $k$ is a scalar multiple of the reproducing kernel of $\B(E_{\H,\ell})$ at the origin. By formula \eqref{eq43}, we have 
$k = -\frac{\mu(\{0\})}{\pi} \cdot \frac{\Theta^{-}_{\H}(\ell, z)}{z}$. From here we see that 
$\supp\mu \subset Z(\Theta_{\H}^-)$, where $Z(\Theta_{\H}^-) \subset \R$ denotes the zero set of the entire function $\Theta_{\H}^{-}(\ell, z)$. Suppose that there exists a point $w \in Z(\Theta_{\H}^-) \setminus \supp\mu$. Then the function 
$g = \Theta_{\H}^{-}(\ell,z)/(z - w)$ belongs to $\B(E_{\H,\ell})$ and $\Pi_\mu g = 0$. The latter contradicts to the unitarity of $\Pi_{\mu}$. Hence,  
$\supp\mu = Z(\Theta_{\H}^-)$. Next, for every $t \in \supp\mu$ we have 
$$
(k_t^\ell(t))^{2}\mu(\{t\}) = \|\Pi_{\mu}k_t^\ell\|_{L^2(\mu)}^{2} = \|k_t^\ell\|_{\B(E_{\H,\ell})}^{2} =k_t^\ell(t),
$$
where $k_t^\ell$ denotes the reproducing kernel of $\B(E_{\H,\ell})$ at the point $t$. This shows that $\mu(\{t\}) = \|k_t^\ell\|_{\B(E_{\H,\ell})}^{-2}$ and hence $\mu$ coincides with the principal spectral measure, see formula \eqref{eq72}. \qed

\medskip

\noindent{\bf Proof of Theorem \ref{t4}.} Let us check that assertion $(a)$ yields assertion $(b)$. As in the proof of Lemma \ref{l10}, consider the Hilbert space 
$$
\B = \Bigl(\pw_{a}, \;\; (f,g)_{\B} = (f,g)_{L^2(\mu)} \Bigr).
$$
Then $\B\sets\pw_{a}$ and the embedding $\Pi_{\mu}: \B \to L^2(\mu)$ is a unitary operator. 
By Theorem \ref{A1}, $\B$ is the de Branges space generated by an Hermite-Biehler function $E$; we may choose $E$ so that $E(0) = 1$. The space $\B$ is regular since so is $ \pw_a $. The function $E$ does not vanish on the real line, for otherwise there would exist a point $w_0$ such that $f(w_0) = 0$ for all $f$ in $\B\sets\pw_a$, see 
\eqref{eq43}. By Theorem \ref{A3}, there exist a number $\ell > 0$ and a Hamiltonian $\H$ on $[0,\ell]$ compatible at $0$ such that $\trace\H(r) = 2$ for almost all $r \in [0,\ell]$ and $E = E_{\H,\ell}$, where $E_{\H,\ell}$ is defined in \eqref{eq82}.  By Lemma \ref{l3} the measure $ \mu $ coincides with the corresponding principal spectral measure.
By Lemma \ref{l10}, the Hamiltonian $\H$ has no indivisible intervals, hence the domain of definition of the Weyl-Titchmarsh transform associated to $\H$ is the whole space~$L^2(\H, \ell)$.

\medskip

Let $\W_{\H_0}: L^2(\H_0, a) \to \pw_a$ and $\W_{\H}: L^2(\H, \ell) \to \B(E_{\H,\ell})$ denote the Weyl-Titchmarsh transforms associated with Hamiltonians $\H_0$ and $\H$, respectively. Consider solutions $\Theta_{\H_0}(\cdot, z)$, $\Theta_{\H}(\cdot, z)$ of Cauchy problem \eqref{eq73} for $\H_0$, $\H$. For every $w \in \C$ formula \eqref{eq43} says 
\begin{equation}\label{eq102}
\W_{\H_0} \Theta_{\H_0}(\cdot, \bar w) = \sqrt{\pi}\cdot k^{\H_0}_{w},
\qquad 
\W_{\H} \Theta_{\H}(\cdot, \bar w) = \sqrt{\pi}\cdot k^{\H}_{w},
\end{equation}
where $k^{\H_0}_{w}$ is the reproducing kernel of $\pw_a$ at the point $w$; $k^{\H}_{w}$ is the reproducing kernel of $\B(E_{\H,\ell})$ at $w$. Define the operator $T_{\mu, a}$ by formula \eqref{eq75} with $s = a$. We claim that $ T_{\mu,a}^{-1} k^{\H_0}_{w} = k^{\H}_{w}$. Indeed, for every function $f \in \pw_a$ we have
$$
f(w) = (f, k^{\H_0}_{w})_{L^2(\R)} = (T_{\mu,a} f, T_{\mu,a}^{-1} k^{\H_0}_{w})_{L^2(\R)} 
= (f, T_{\mu,a}^{-1} k^{\H_0}_{w})_{L^2(\mu)}. 
$$
This formula and the unitarity of the embedding $\Pi_\mu$ means that $T_{\mu,a}^{-1} k^{\H_0}_{w}$ regarded as an element of $\B(E_{\H,\ell})$ coincides with $k^{\H}_{w}$, that is, 
$ T_{\mu,a}^{-1} k^{\H_0}_{w} = k^{\H}_{w}$. Finally, the operator 
$\W_{\H}^{-1} T^{-1}_{\mu, a}\W_{\H_0}: L^2(\H_0, a) \to L^2(\H, \ell)$ is bounded, invertible and coincides with $S_\H$ on complete family of functions $\Theta_{\H_0}(\cdot, \bar w)$, $w \in \C$, which yields assertion~$(b)$. 

\medskip
  
Now let assertion $(b)$ be satisfied. Then the Hamiltonian $\H$ is compatible at~$0$ and $\ell$, and $\Pi_\mu: \B(E_{\H,\ell}) \to L^2(\mu)$ is a unitary operator, see Section \ref{s2}. Let us keep notation $S_\H$ for the extension to the whole space $L^2(\H_0, a)$ of the operator $S_\H$ in the statement of Theorem \ref{t4}. By assumption, $S_\H: L^2(\H_0, a) \to L^2(\H, \ell)$ is bounded and invertible. Consider the operator $T = \W_{\H} S_\H \W^{-1}_{\H_0}$ from $\pw_a$ to $\B(E_{\H,\ell})$. Observe that formula \eqref{eq102} does not depend of assumption $(a)$, hence  
\begin{equation}\label{eq103}
T k^{\H_0}_{w} = k^{\H}_{w}, \qquad w \in \C. 
\end{equation}
Define the Hilbert space 
$$
\tilde\B = \Bigl(\pw_{a}, \;\; (f,g)_{\tilde\B} = (T^{-1}(T^{-1})^{\ast}f,g)_{L^2(\R)} \Bigr).
$$
Since $T$ is bounded and invertible, we have $\tilde\B \sets \pw_a$. Next, for every $f \in \pw_a$ and  $w \in \C$ we have
\begin{equation}\notag
f(w) 
= (f, k_{w}^{\H_0})_{L^2(\R)} = (T^{-1}(T^{-1})^{\ast} f, T^\ast T k_{w}^{\H_0})_{L^2(\R)} 
= (f, T^\ast T k_{w}^{\H_0})_{\tilde\B}.
\end{equation}
Thus, the function $T^\ast T k_{w}^{\H_0} \in \pw_a$ regarded as an element of $\tilde\B$ is the reproducing kernel at the point $w$. On the other hand, for all $w \in \C$ we have
\begin{equation}\notag
\begin{aligned}
(T^\ast T k_{w}^{\H_0})(z) &= (T^\ast T k_{w}^{\H_0}, k_{z}^{\H_0})_{L^2(\R)} = (T k_{w}^{\H_0}, Tk_{z}^{\H_0})_{\B(E_{\H,\ell})} \\
&= (k_{w}^{\H}, k_{z}^{\H})_{\B(E_{\H,\ell})} = k_{w}^{\H}(z),
\end{aligned}
\end{equation}
where we used formula \eqref{eq103}. This shows that Hilbert spaces of entire functions $\tilde \B$ and $\B(E_{\H,\ell})$ have the same reproducing kernels. Hence, they coincide as Hilbert spaces. It follows that 
$\B(E_{\H, \ell}) = \tilde\B \sets \pw_a$. Now assertion $(a)$ follows from the fact that the embedding operator $\Pi_\mu: \B(E_{\H,\ell}) \to L^2(\mu)$ is unitary.  \qed

\medskip

The proof of the following elementary proposition is left to the reader.

\begin{Prop}\label{p1}
Let $\H$ be a regular Hamiltonian on $[0, \ell]$ having no indivisible intervals. The operator $S_{\H}$ in (\ref{SH}) can be extended to the whole space $L^2(\H_0,a)$ as a bounded and boundedly invertible operator from $L^2(\H_0,a)$ onto $L^2(\H,\ell)$ if and only if the same is true of the operator 
$\tilde S_\H: L^2(\H_0, a) \to L^2(\H, \ell)$ densely defined by
\begin{equation}\label{eq104}
\tilde S_\H : {J^\ast}^{n} 
\left(\!
\begin{smallmatrix}
\frac{t^n}{n!}\\
0\\
\end{smallmatrix}
\!\right) 
\mapsto 
\int_{0}^{t}\!J^*\H(t_1)\int_{0}^{t_1}\!J^*\H(t_2)\ldots\int_{0}^{t_{n-1}}\!J^*\H(t_n)\oz d t_n \ldots d t_1, \; n \ge 0 .
\end{equation}
Moreover, their extensions coincide if they exist. 
\end{Prop}

\noindent{\bf Hint.} Let $\partial^{n}_{0}\Theta_{\H_0}(\cdot, 0)$ denote the derivative with respect to $w$ of order $n$ at $w = 0$ of the analytic mapping $w \mapsto \Theta_{\H_0}(\cdot, w)$ from $\C$ to $L^2(\H_0,a)$, and let $\partial^{n}_{0}\Theta_{\H}(\cdot, 0)$ be defined analogously. Then 
$S_\H \partial^{n}_{0}\Theta_{\H_0}(\cdot, 0) = \partial^{n}_{0}\Theta_{\H}(\cdot, 0)$. On the other hand, we have
$$
\frac{1}{n!}\cdot \partial^{n}_{0}\Theta_{\H}(\cdot, 0) =
\int_{0}^{t}\!J^*\H(t_1)\int_{0}^{t_1}\!J^*\H(t_2)\ldots\int_{0}^{t_{n-1}}\!J^*\H(t_n)\oz d t_n \ldots d t_1 
$$
and similarly, $\frac{1}{n!}\partial^{n}_{0}\Theta_{\H_0}(\cdot, 0) = 
{J^\ast}^{n}\left(\!
\begin{smallmatrix}
\frac{t^n}{n!}\\
0\\
\end{smallmatrix}
\!\right)
$. This is enough for the proof of Proposition.

\medskip

As a corollary, we obtain the following necessary condition in the diagonal case.
\begin{Prop}\label{p2}
Let $\H$ be a Hamiltonian on $[0,a]$ of the form 
$
\H = 
\left(
\begin{smallmatrix}
w & 0 \\
0 &1/w
\end{smallmatrix}
\right)
$
with some function $w$ such that $w(s)>0$ for almost all $s \in [0,a]$. Assume that equivalent assertions in Theorem \ref{t4} are satisfied. Put $\phi = \log w$. Then there exist positive constants $c_1$, $c_2$ such that
$$
 c_1 \le \frac{1}{a_n(s)}\int_{0}^{s}e^{(-1)^n \phi(t)}\left(\int_{0}^{t}\!\!\int_{0}^{t_1}\!\!\dots\int_{0}^{t_{n-1}} e^{\sum\nolimits_{1}^{n}(-1)^{n+k}\phi(t_k)}\,dt_n\ldots dt_1\right)^2 \!\! dt \le c_2,
$$
for all positive $s \le a$ and all integer $n \ge 1$, where $a_n(s) = \frac{s^{2n+1}}{n\cdot (n!)^2}$.
\end{Prop}
\noindent{\bf Hint.} By Krein's formula, we have $\B(E_{\H,s}) = \pw_s$ for all $s \in [0,a]$, see Remark after Lemma \ref{l10}. Therefore, for all integers $n \ge 1$ the $L^2(\H_0, s)$-norm of the left hand side of~\eqref{eq104} is comparable to the $L^2(\H, s)$-norm of the right hand side of~\eqref{eq104}. 

\section{Proof of Theorem \ref{t1}}\label{s3}
We start with the following simple lemma. 
\begin{Lem}\label{l12}
Let $\H$ be a regular Hamiltonian on $[0, \ell]$ having no indivisible intervals, and let $\Theta_\H$ be the solution of Cauchy problem \eqref{eq73}. Then for all $r \in [0, \ell]$ we have
\begin{align}
\int_{0}^{r}\bigr\langle\H(t)\oz,\oz\bigr\rangle_{\C^2}\, dt 
&=\frac{1}{\pi}\left\|\frac{\Theta^{-}_{\H}(r,t)}{t}\right\|^{2}_{L^2(\mu)}\label{eq161}, \\
\int_{0}^{r}\bigr\langle\H(t)\zo,\zo\bigr\rangle_{\C^2}\, dt 
&= \frac{1}{\pi}\left\|\frac{\Theta^{+}_{\H}(r,t) - 1}{t}\right\|^{2}_{L^2(\mu)} \label{eq162}, \\
\int_{0}^{r}\bigr\langle\H(t)\oz, \zo\bigr\rangle_{\C^2}\, dt 
&= -\frac{1}{\pi}\left(\frac{\Theta^{-}_{\H}(r,t)}{t}, \frac{\Theta^{+}_{\H}(r,t) - 1}{t}\right)_{L^2(\mu)} \label{eq163}, 
\end{align}
where $\mu$ is the principal spectral measure from (\ref{eq47}).
\end{Lem}
\beginpf Take $r \in [0,\ell]$. Since the Hamiltonian $\H$ has no indivisible intervals, the operators $\W_r: L^2(\H, r) \to \B(E_{\H,r})$ and $\Pi_{\mu}: \B(E_{\H,\ell}) \to L^2(\mu)$ defined in \eqref{eq77}, \eqref{eq114} are unitary and the operator $\Pi_\mu\W_r: L^2(\H, r) \to L^2(\mu)$ is isometric. We claim that
$$
\W_r \oz = -\frac{1}{\sqrt{\pi}}\cdot \frac{\Theta^{-}_{\H}(r,z)}{z}, \qquad  
\W_r \zo = \frac{1}{\sqrt{\pi}} \cdot \frac{\Theta^{+}_{\H}(r,z) - 1}{z}.
$$
Indeed, the first formula above is \eqref{eq43} for $w = 0$. The second formula can be obtained from the following computation:
$$
z\int_{0}^{r} \bigl\langle H(t)\zo, \Theta_{\H}(t,\bar z)\bigr\rangle_{\C^2}\,dt = 
\int_{0}^{r} \bigl\langle \zo, J\partial_t\Theta_{\H}(t,\bar z)\bigr\rangle_{\C^2}\,dt
= \Theta_{\H}^+(r,z) - 1.
$$
It remains to use the fact that $\Pi_\mu\W_r$ is an isometry. \qed

\medskip

The main observation allowing to solve the inverse problem is formulated as follows. 

\begin{Lem}\label{l42}
Let $\mu$ be the measure such that the assumptions $(1)-(3)$ are satisfied and let $\H(r)$, $r \in [0 ,\ell]$, be the corresponding Hamiltonian, $\{\B(E_{\H, r})\} $ be its be Branges chain, $\xi(s) $ be the function defined in Lemma \ref{l10}. Define $k^{r}_{w}$ to be the reproducing kernel of the space $\B(E_{\H,r}) $.  Then 
\[T_{\mu, s}^{-1} \sinc_s(t- \bar w) = k^{\xi (s)}_w, \qquad w \in \C, \quad s \in [0, a], \]
where $\sinc_s(t -\bar w) = \frac{\sin s ( t -\bar w ) }{ \pi(t-\bar w)}  $ is the reproducing kernel of the space $ \pw_s$.
\end{Lem}

\beginpf 
For any $f \in \pw_s$ we have
\begin{multline*}
\left(f, k^{\xi(s)}_{w} \right)_{\B(E_{\H, \xi(s)})} 
=  f(w) = \left(f, \sinc_s(t -\bar w)\right)_{L^2(\R)} =\\
= \left(f, T_{\mu,s}^{-1}\sinc_s(t -\bar w)\right)_{L^2(\mu)}
= \left(f, T_{\mu,s}^{-1}\sinc_s(t -\bar w)\right)_{\B(E_{\H, \xi(s)})},
\end{multline*}
as required.
\qed

\medskip

Given an entire function $F$ and a point $z \in \C$, we will denote by $\dot{F}(z)$, $\ddot{F}(z)$ the values of its first and second derivatives at the point $z$, correspondingly.

\begin{Lem}\label{l14}
Let $\H$, $\Theta_{\H}$, and $\mu$ be as in Lemma \ref{l12}. Then 
$\mu(\{0\}) = -1/\dot{\Theta}_{\H}(\ell,0)$,
$$
\Theta^{+}_{\H}(\ell, t) = -\frac{\pi}{\mu(\{t\})\dot{\Theta}_{\H}^{-}(\ell, t)}, \quad \dot{\Theta}^{+}_{\H}(\ell,0) = \frac{\hat{K}_{\mu}(0) + c}{\mu(\{0\})}+ \frac{1}{2}\ddot{\Theta}^{-}_{\H}(\ell, 0)\mu(\{0\}),
$$
where $t\in \supp\mu\setminus\{0\}$, $c$ is the constant from \eqref{eq47} and $\hat{K}_{\mu}(0) = \frac{1}{\pi}\int_{\R\setminus\{0\}}\frac{d\mu(t)}{t(1+t^2)}$.
\end{Lem}
\beginpf Let $M$ be the fundamental matrix solution \eqref{eq66} of system \eqref{eq1}. Then $\Theta_\H$ is the first column of $M$. Since $\det M(r, z) = 1$ for all $r \in [0, \ell]$ and all $z \in \C$, we have
$$
\Theta_\H^+(\ell,z)\Phi_\H^-(\ell, z) - \Theta_\H^-(\ell,z)\Phi_\H^+(\ell, z) = 1, \qquad z \in \C.
$$
In particular, $\Theta_\H^+(\ell,t) = 1/\Phi_\H^-(\ell, t)$ for all $t \in \supp\mu$ (see formula \eqref{eq72}) and we have $\dot{\Theta}_{\H}^{+}(\ell, 0) = - \dot{\Phi}_{\H}^{-}(\ell, 0)$. Now the statement follows from \eqref{eq47} using a straightforward residue calculus. \qed

\medskip

\noindent{\bf Proof of Theorem \ref{t1}.} Let $\mu$ be a non-zero measure with properties $(1) -(3)$. Then there exists a Hamiltonian $\H$ on an interval $[0,\ell]$ such that $\H$ has no indivisible intervals, $\trace\H(r) = 2$ for almost all $r \in [0, \ell]$, $\mu$ is the corresponding principal spectral measure, and $\B(E_{\H, \ell}) \sets \pw_a$, see the proof of Theorem \ref{t4}. By Lemma \ref{l10}, there is an increasing bijection $\xi: [0, a] \to [0, \ell]$ such that $\B(E_{\H, \xi(s)}) \sets \pw_s$ for all $s \in [0, a]$. Fix a number $s \in (0, \ell]$. We claim that 
\begin{equation}\label{eq109}
G_{1, s}(t) = -\frac{\Theta^{-}_{\H}(\xi(s),t)}{t}, \quad G_{2, s}(t) = \frac{\Theta^{+}_{\H}(\xi(s),t) - 1}{t}, 
\qquad t \in \supp \mu.
\end{equation}
Applying Lemma \ref{l42} with $ w = 0 $ we find that 
 
$$
-\frac{\Theta^{-}_{\H}(\xi(s),z)}{\pi z} =  \left(T_{\mu,s}^{-1}\frac{\sin sx}{\pi x}\right)(z) = 
\left(T_{\mu,s}^{-1}\frac{\sin (sx)}{\pi x}, \frac{\sin s(x - \bar z)}{\pi(x - \bar z)}\right)_{L^2(\R)}.
$$
This yields the the first formula in \eqref{eq109} and gives 
$$G'_{1,a}(0) = -\frac{1}{2}\ddot{\Theta}^{-}_{\H}(\ell,0), \qquad G'_{1,a}(t) = - \frac{\dot{\Theta}^{-}_{\H}(\ell,t)}{t}, 
\quad t \in \supp\mu\setminus\{0\}.$$
To check the second formula in \eqref{eq109}, we first consider the case where $s = a$. 
We have
\begin{equation*}
G_{2,a}(0) = \frac{\hat{K}_{\mu}(0) + c}{\mu(\{0\})}- G'_{1,a}(0)\mu(\{0\}) 
= \frac{\hat{K}_{\mu}(0) + c}{\mu(\{0\})}+ \frac{1}{2}\ddot{\Theta}^{-}_{\H}(\ell, 0)\mu(\{0\})
\end{equation*}
for $t=0$, and 
\begin{equation*}
G_{2,a}(t) = \frac{1}{t}\left(\frac{\pi}{t \mu(\{t\}) G'_{1,a}(t)} - 1 \right) = 
\frac{1}{t}\left(-\frac{\pi}{\mu(\{t\}) \dot{\Theta}^{-}_{\H}(\ell,t)} - 1 \right)
\end{equation*}
for $ t \in\supp\mu\setminus\{0\}$. Applying Lemma \ref{l14}, we see that $G_{2, a}(t) = \frac{\Theta^{+}_{\H}(\ell,t) - 1}{t}$ for all points $t \in \supp\mu$. Now take $0<s<a$ and denote by $P_{\xi(s)}$ the orthogonal projection in $L^2(\H, \ell)$ to $L^2(\H, \xi(s))$ and by $\mathcal{P}_{\xi(s)}$ the orthogonal projection in $\B(E_{\H, \ell})$ to $\B(E_{\H, \xi(s)})$. Let $\W_{\ell}$, $\W_{\xi(s)}$ be the Weyl-Titchmarsh transforms associated with the Hamiltonian $\H$ on $[0,\ell]$ and $[0, \xi(s)]$. Note that $\W_{\ell}P_{\xi(s)} = 
\mathcal{P}_{\xi(s)} \W_{\ell}$. For every point $z \in \C$ we have:
\begin{align*}
\frac{\Theta^{+}_{\H}(\xi(s),z) - 1}{z} 
&=\sqrt{\pi}\bigl(\W_{\xi(s)}\!\zo\bigr)(z) = \sqrt{\pi}\bigl(\W_{\ell}P_{\xi(s)}\!\zo\bigr)(z) 
= \sqrt{\pi}\bigl(\mathcal{P}_{\xi(s)} \W_{\ell}\!\zo\bigr)(z) \\
&= \left(\mathcal{P}_{\xi(s)}\frac{\Theta^{+}_{\H}(\ell,x) - 1}{x}\right)(z) = \left(\frac{\Theta^{+}_{\H}(\ell, x) - 1}{x}, k^{\xi(s)}_{z}\right)_{\B(E_{\H, \ell})}\\
&= \left(\frac{\Theta^{+}_{\H}(\ell,x) - 1}{x},\; T_{\mu,s}^{-1}\frac{\sin s(x - \bar z)}{\pi(x - \bar z)}\right)_{\B(E_{\H, \ell})} \\
&= \left(G_{1, a},\; T_{\mu,s}^{-1}\frac{\sin s(x - \bar z)}{\pi(x - \bar z)}\right)_{L^2(\mu)}.
\end{align*}
It follows that $G_{2,s}(t) = \frac{\Theta^{+}_{\H}(\xi(s),t) - 1}{t}$ for all $s \in [0,a]$ and all $t \in \supp\mu$. 
To complete the proof of the Theorem, use Lemma \ref{l12} and compute
\begin{align*}
2\xi(s) 
&= \int_{0}^{\xi(s)}\trace\H(t)\,dt =  \frac{1}{\pi}\left\|\frac{\Theta^{-}_{\H}(\xi(s),t)}{t}\right\|^{2}_{L^2(\mu)} + 
\frac{1}{\pi}\left\|\frac{\Theta^{+}_{\H}(\xi(s),t) - 1}{t}\right\|^{2}_{L^2(\mu)}\\
&= \frac{1}{\pi}\|G_{1, s}\|_{L^2(\mu)}^{2} + \frac{1}{\pi}\|G_{2, s}\|_{L^2(\mu)}^{2} = G_{1, s}(0) + \frac{1}{\pi}\|G_{2, s}\|_{L^2(\mu)}^{2}.
\end{align*}
Thus, the function $\zeta$ defined in Section \ref{s1} coincides with the function $\xi$. Using Lemma \ref{l12} again, we obtain 
\begin{align*}
\int_{0}^{\xi(s)}\bigr\langle\H(t)\oz,\oz\bigr\rangle_{\C^2}\,dt
&= \frac{1}{\pi} \|G_{1, s}\|^{2}_{L^2(\mu)} =  G_{1, s}(0),\\
\int_{0}^{\xi(s)}\bigr\langle\H(t)\oz,\zo\bigr\rangle_{\C^2}\,dt 
&= \frac{1}{\pi} (G_{1, s}, G_{2, s})_{L^2(\mu)}.
\end{align*}
Since $\trace\H(t)=2$ for almost all $t \in [0, \ell]$, we also have
$$
\int_{0}^{\xi(s)}\bigr\langle\H(t)\zo,\zo\bigr\rangle_{\C^2}\,dt 
 = 2\xi(s) - G_{1,s}(0).\\
$$
Let $\tau: [0, \ell] \to [0, a]$ be the inverse function to $\zeta = \xi$ and let the Hamiltonian $\H$ be of the form 
$\H = 
\left(
\begin{smallmatrix}
h_1 & h \\
h & h_2
\end{smallmatrix}
\right)
$. Then 
$$
\int_{0}^{r} h_1(t)\,dt = G_{1, \tau(r)}(0),  
\int_{0}^{r} h_2(t)\,dt = s - G_{1, \tau(r)}(0),
$$ 
$$
\int_{0}^{r} h_1(t)\,dt = \frac{1}{\pi}(G_{1, \tau(r)}, G_{2, \tau(r)})_{L^2(\mu)}
$$
for all $r \in [0, \ell]$. In particular, the functions $g_1$, $g_2$, $g$ defined in \eqref{eq111} are absolutely continuous and the Hamiltonian in the right hand side of formula \eqref{eq48} coincides with $\H$ almost everywhere on $[0, \ell]$. \qed  

\section{Example of non-Paley-Wiener Hamiltonian}

The following example shows that a Hamiltonian $ \H $ on $ [ 0 , \ell ] $ can be bounded away from $0 $ and $ \infty $ but the space $ \B ( E_{ \H , \ell }) $ be not equivalent to a Paley--Wiener space.  

\medskip

The Hamiltonian in question will be diagonal and take two values. Let $I_j$, $j\ge 1$, be consecutive subintervals of $(0,1/2)$  of length $l_j = 3^{-j}$ accumulating at $1/2$ (the left end of $I_1$ is zero). Given an $h\in(0,1) $ define
$$
\H(x) = \begin{cases} 
\idm, 
& \!\! x \in \bigcup_{j\ge1} I_{2j-1} \cr
\left(\begin{smallmatrix} 
h & 0 \\ 0 & h^{ -1 }  
\end{smallmatrix}\right) , & \!\! x \in \bigcup_{j\ge 1} I_{2j}. 
\end{cases} 
$$ 

\begin{Prop}\label{separated}
For any $  h < 1/9 $ the space $\B ( E_{\H, 1/2 } ) \not\sets \pw_a $ for any $ a > 0 $.
\end{Prop} 

\beginpf 
Let $E = E_{\H,1/2}$. It suffices to show that $E(\lambda) \neq O (|\lambda|)$ on the real axis as 
$\lambda\to\infty$. Indeed, if $\B(E)\sets\pw_a$, then the function $E(z)/(z-z_0)$ belongs to 
$\pw_a$ for any $z_0$ such that $E(z_0)=0$, and hence is bounded on $\R$.

Let $M(\lambda) = M (1/2, \lambda)$ be the fundamental matrix for the system with the Hamiltonian $\H(x)$, 
$\Theta$ the first column of $M$. By the chain rule we have
\begin{align*}
M(\lambda) &= T_n(\lambda) M_n(\lambda) M_{n-1} (\lambda) \cdots M_2 (\lambda) M_1 (\lambda), \\
T_n(\lambda) &= \cdots M_{n+2}(\lambda) M_{n+1}(\lambda), \\
M_j(\lambda) &= 
\begin{cases} 
\left(\!
\begin{smallmatrix}
\cos\lambda l_j & \sin\lambda l_j \\  
-\sin \lambda l_j & \cos \lambda l_j 
\end{smallmatrix}
\!\right), 
&j\,\mbox{ odd}, \\
\left(\!
\begin{smallmatrix}
\cos \lambda l_j & h^{-1} \sin \lambda l_j \\
-h \sin \lambda l_j & \cos \lambda l_j 
\end{smallmatrix}
\!\right), 
&j\,\mbox{ even}.
\end{cases} 
\end{align*}
Let $\lambda_k = \pi 3^k/2$, $k$ being even. Then for $j \le k$
$$
M_j(\lambda_k) = 
\begin{cases} 
\left(\!
\begin{smallmatrix}
0 & \pm 1 \\
\mp 1 & 0 
\end{smallmatrix}
\!\right), &j\,\mbox{ odd}, \\
\left(\!
\begin{smallmatrix} 
0 & \pm h^{ -1 } \\
\mp h & 0 
\end{smallmatrix}
\!\right), &  j\,\mbox{ even}.
\end{cases} 
$$
The signs here depend on the oddity of $j$ and are of no matter for us. It follows that 
$$
M_k(\lambda_k) M_{k-1} (\lambda_k) \cdots M_2 (\lambda_k) M_1 (\lambda_k) =  (-1)^{k/2} 
\left(\!
\begin{smallmatrix}
h^{ - k/2 }& 0  \\ 
0 & h^{ k/2 } 
\end{smallmatrix}
\!\right).
$$
Thus, $\Theta(\lambda_k) = T_k(\lambda_k) 
\left(\!
\begin{smallmatrix}
h^{-k/2} \\ 
0 
\end{smallmatrix}
\!\right)$. 
We  estimate
\begin{equation}\label{q} 
h^{-k/2} \le \bigl\|T_k\left(\lambda_k\right)^{-1}\bigr\|\cdot\bigl\|\Theta(\lambda_k)\bigr\|. 
\end{equation}
The norm of each factor of $T_k(\lambda)$ is estimated above as $\|M_j(\lambda)\| \le 1 + h^{-1} |\lambda| l_j$, therefore 
\[ 
\bigl\| T_k \left(\lambda_k\right)^{-1} \bigr\| \le \prod_{j > k} \|M_j^{-1}(\lambda_k)\| 
= \prod_{j > k} \|M_j(\lambda_k)\|\le\prod_{j>k} \left(1 + h^{-1} 3^{k-j}\right),
\] 
the right hand side being a constant in $ k $. Plugging this in (\ref{q}) we find that
\[|E(\lambda_k)|=\left\|\Theta(\lambda_k)\right\|\ge Ch^{-k/2}. 
\] 
Taking $h\in(0,1/9)$ we obtain the required assertion.     
\qed

\bibliographystyle{plain} 
\bibliography{bibfile}

\def\cprime{$'$} \def\cprime{$'$} \def\cprime{$'$}
\begin{thebibliography}{10}

\bibitem{AlbeverioGrinivMykytyuk}
S.~Albeverio, R.~Hryniv, and Ya. Mykytyuk.
\newblock Inverse spectral problems for {D}irac operators with summable
  potentials.
\newblock {\em Russ. J. Math. Phys.}, 12(4):406--423, 2005.

\bibitem{BBK}
Anton Baranov, Roman Bessonov, and Vladimir Kapustin.
\newblock Symbols of truncated {T}oeplitz operators.
\newblock {\em J. Funct. Anal.}, 261(12):3437--3456, 2011.

\bibitem{dbbook}
Louis de~Branges.
\newblock {\em Hilbert spaces of entire functions}.
\newblock Prentice-Hall, Inc., Englewood Cliffs, N.J., 1968.

\bibitem{Den06}
Sergey~A. Denisov.
\newblock Continuous analogs of polynomials orthogonal on the unit circle and
  {K}re\u\i n systems.
\newblock {\em IMRS Int. Math. Res. Surv.}, pages Art. ID 54517, 148, 2006.

\bibitem{EckhardtKostenkoTeschl}
Jonathan {Eckhardt}, Aleksey {Kostenko}, and Gerald {Teschl}.
\newblock {Inverse uniqueness results for one-dimensional weighted Dirac
  operators.}
\newblock In {\em {Spectral theory and differential equations: V. A.
  Marchenko's 90th anniversary collection}}, pages 117--133. Providence, RI:
  American Mathematical Society (AMS);, 2014.

\bibitem{HWdeSnoo}
Seppo Hassi, Henk De~Snoo, and Henrik Winkler.
\newblock Boundary-value problems for two-dimensional canonical systems.
\newblock {\em Integral Equations Operator Theory}, 36(4):445--479, 2000.

\bibitem{HrynivMykytyuk2003}
R.~O. Hryniv and Ya.~V. Mykytyuk.
\newblock Inverse spectral problems for {S}turm-{L}iouville operators with
  singular potentials.
\newblock {\em Inverse Problems}, 19(3):665--684, 2003.

\bibitem{HrynivMykytyuk2004}
Rostyslav~O. Hryniv and Yaroslav~V. Mykytyuk.
\newblock Transformation operators for {S}turm-{L}iouville operators with
  singular potentials.
\newblock {\em Math. Phys. Anal. Geom.}, 7(2):119--149, 2004.

\bibitem{Levinlectures}
B.~Ya. Levin.
\newblock {\em Lectures on entire functions}, volume 150 of {\em Translations
  of Mathematical Monographs}.
\newblock American Mathematical Society, Providence, RI, 1996.
\newblock In collaboration with and with a preface by Yu. Lyubarskii, M. Sodin
  and V. Tkachenko.

\bibitem{LSbook}
B.~M. Levitan and I.~S. Sargsjan.
\newblock {\em Sturm-{L}iouville and {D}irac operators}, volume~59 of {\em
  Mathematics and its Applications (Soviet Series)}.
\newblock Kluwer Academic Publishers Group, Dordrecht, 1991.
\newblock Translated from the Russian.

\bibitem{LS}
Yurii~I. Lyubarskii and Kristian Seip.
\newblock Weighted {P}aley-{W}iener spaces.
\newblock {\em J. Amer. Math. Soc.}, 15(4):979--1006 (electronic), 2002.

\bibitem{Mar06}
V.~A. Marchenko.
\newblock The generalized shift, transformation operators, and inverse
  problems.
\newblock In {\em Mathematical events of the twentieth century}, pages
  145--162. Springer, Berlin, 2006.

\bibitem{Remling}
Christian Remling.
\newblock Schr{\"o}dinger operators and canonical systems.
\newblock In {\em Operator Theory}, pages 1--7. Springer Basel, 2015.

\bibitem{Romanov}
Roman Romanov.
\newblock Canonical systems and de {B}ranges spaces.
\newblock {\em preprint arXiv:1408.6022}, 2014.

\bibitem{Sar07}
Donald Sarason.
\newblock Algebraic properties of truncated {T}oeplitz operators.
\newblock {\em Oper. Matrices}, 1(4):491--526, 2007.

\bibitem{Volberg}
A.~L. Vol{\cprime}berg.
\newblock Thin and thick families of rational fractions.
\newblock In {\em Complex analysis and spectral theory ({L}eningrad,
  1979/1980)}, volume 864 of {\em Lecture Notes in Math.}, pages 440--480.
  Springer, Berlin-New York, 1981.

\end{thebibliography}
\enddocument